\documentstyle[aps]{revtex}
\begin{document}
\draft
\tighten
\title{RELATIVISTIC ANTIHYDROGEN PRODUCTION}
\author{Helmar Meier$^1$, Zlatko Halabuka$^1$, Kai Hencken$^1$,
	 Dirk Trautmann$^1$ and Gerhard Baur$^2$}
\address{$^1$Institut f\"ur Physik der Universit\"at Basel,
          4056 Basel, Switzerland \\
          $^2$Institut f\"ur Kernphysik (Theorie), Forschungszentrum J\"ulich, 
          52425 J\"ulich, Germany}
\date{\today}
\maketitle
\begin{abstract}
Antihydrogen has recently been produced in collisions of 
antiprotons with ions. While passing through the Coulomb 
field of a nucleus an antiproton will create an 
electron-positron pair. In rare cases the positron is bound 
by the antiproton and an antihydrogen atom produced.
We calculate the production of relativistic antihydrogen 
atoms by bound-free pair production. The cross section is 
calculated in the semiclassical approximation (SCA), or 
equivalently in the plane wave Born approximation (PWBA)
using exact Dirac-Coulomb wave functions. 
We compare our calculations to the equivalent photon 
approximation (EPA).
\end{abstract}
\section{Introduction}
Antihydrogen, the simplest bound state of antimatter, has first been produced 
and detected in 1995 at CERN in the 
Low Energy Antiproton Ring (LEAR) \cite{cern}.
The synthesis of an antiproton and a positron has been done by passing 
relativistic antiprotons through a xenon target. Only a few electron-positron 
pairs are produced in these collisions. 
In rare cases the velocities of the outgoing 
positron and antiproton are sufficiently close together, 
so that the particles join.
In lowest order the formalism for calculating the production of 
antihydrogen is the same as the one used to study
bound-free pair production in ion-ion collisions. 
Bound-free pair production is an important 
process as it is one of the reactions 
limiting the luminosity of heavy ion beams in high energy 
colliders \cite{datz}.
The equivalent-photon approach (Weizs\"acker-Williams method) 
has been used for the calculations of antihydrogen in \cite{brodsky} 
and for electron capture by a heavy ions in \cite{aste,agger}.
It is of interest to perform exact calculations in the framework of SCA
or, equivalently, PWBA theory
and compare them with these calculations, especially at low energies
of the colliding particle. 
The bound-free pair production by bremsstrahlung has been considered in 
\cite{baur} and was found to be negligible. A recent calculation of Baltz 
shows that contributions of higher order effects reduce the cross section 
given in lowest order perturbation theory by a small amount for
$\gamma\rightarrow\infty$ \cite{baltz}. 
\section{Total cross section}
The total SCA cross section for 
pair production with electron capture in a heavy ion collision 
is given by \cite{eichler}
\begin{equation}
\sigma_{tot} = 8 \pi \left(\frac{Z_P \alpha}{\beta} \right)^2
\int \limits_{m_e}^\infty dE_i \int \limits_{q_z}^\infty
ds\frac{s}{[s^2-(\beta q_z)^2]^2}\sum_{\kappa_i}
\sum_{m_i,m_f}\left|\left\langle \psi_f(\vec r \,)
\left|(1-\beta\alpha_3)e^{i\vec q\vec r}\right|\psi_i(\vec r \,)
\right\rangle\right|^2\ \ .
\label{cs}
\end{equation}
Throughout the paper we will set $\hbar=c=1$. The charge numbers of the 
projectile and target are denoted by $Z_P$ and $Z_T$;
the velocity of the projectile in the target rest frame is given by 
$\beta$. 
The momentum transfer from projectile to target is $\vec q$ 
whose absolute value is given by $s=|\vec q\,|$. 
The component of $\vec q$ in the direction of the projectile is 
$q_z=\frac \omega \beta=\frac{E_f-E_i}{\beta}$.
The total energy of the bound electron is $E_f$; 
the one of the positron in the 
continuum $|E_i|$. (Please note that $E_i$ is negative.)
The third component of the Dirac matrices is $\alpha_3$.
In our calculation $\psi_f$ ($=\psi_{\kappa_f}^{m_f}$) is the 
Dirac-Coulomb wave function of a {\em K}-shell electron ($\kappa_f=-1$, 
$E_f=m_e\sqrt{1-\zeta^2}$ with $\zeta=\alpha Z_T$ and magnetic
quantum number $m_f$). $\psi_i$ ($=\psi_{\kappa_i}^{m_i}$) is the wave 
function of an electron with negative energy $E_i$ in the continuum 
describing the positron \cite{eichler}.

Because of charge-conjugation invariance the same formalism is used to 
calculate the production of relativistic antihydrogen in the bound-free 
process: $\overline p + Z_P \rightarrow \overline H(1s)+Z_P+e^-$. 

Using current-conservation we can write (\ref{cs}) as:
\begin{equation}
\sigma_{tot} = 8 \pi \left(\frac{Z_P \alpha}{\beta} \right)^2
\int \limits_{m_e}^\infty dE_i \int \limits_{q_z}^\infty
ds\ s\sum_{\kappa_i}
\sum_{m_i,m_f}\left|\left\langle \psi_f(\vec r \,)
\left|\left(\frac 1 {s^2}-\frac{\vec \beta_{\perp}\vec \alpha}{s^2-(\beta q_z)
^2}\right)
e^{i\vec q\vec r}\right|\psi_i(\vec r \,)
\right\rangle\right|^2\ \ .
\label{cscg}
\end{equation}
It should be mentioned that (\ref{cs}) and (\ref{cscg}) are 
not exactly equal if the wave functions are not exact 
eigenfunctions of the Dirac Hamiltonian \cite{eichler,amundsen}.
The vector $\vec \beta_{\perp}$ is 
perpendicular to $\vec q$ and defined by
\begin{equation}
\vec \beta_{\perp}=\vec \beta - \left(\vec \beta\frac{\vec q} s \right)
\frac{\vec q} s\ \ .
\end{equation} 

The solutions of the Dirac equation in the Coulomb field
\begin{equation} 
V(r)=-\frac{\zeta}{r}\ ,\ \ \ \ \ \ \ \zeta=\alpha Z_T
\label{pot}
\end{equation}
are given by \cite{eichler,rose}
\begin{eqnarray}
\psi_{\kappa}^{m}=\left(
\begin{array}{c}
g_{\kappa}(r)\chi_\kappa^m(\hat r) \\
if_{\kappa}(r)\chi_{-\kappa}^m(\hat r)
\end{array} \right)\ \ .
\label{wf}
\end{eqnarray}
The angular dependence is expressed by the spin-angular functions
\begin{eqnarray}
\chi_\kappa^m(\hat r)=\sum_{\tau=\pm \frac 1 2} 
(-1)^{l+m-\frac 1 2}\,\sqrt{2j+1}\,
\left(
\begin{array}{ccc}
l & \frac 1 2 & j \\
m-\tau & \tau & -m
\end{array}\right)
Y_l^{m-\tau}(\hat r)\,\chi_\tau\ \ .
\label{swf}
\end{eqnarray}
$\chi_\tau$ are Pauli spinors and
\begin{equation}
j=|\kappa|-\frac 1 2\,,\ \ \  l=j+\frac 1 2\mbox{sgn}(\kappa)\ \ .
\end{equation}
The radial functions $g_{\kappa}$ and $f_{\kappa}$ for the bound state are
\begin{eqnarray}
\left(
\begin{array}{c}
g_{-1}(r) \\
f_{-1}(r)
\end{array}
\right)
 = a_0^{-\frac 3 2}\frac{(2Z_T)^{\gamma(-1)+\frac 1 2}}
{[2\Gamma(2\gamma(-1)+1)]^{\frac 1 2}}
\left(
\begin{array}{c}
-(1+\gamma(-1))^{\frac 1 2} \\
(1-\gamma(-1))^{\frac 1 2}
\end{array}\right)
 \left(\frac r {a_0}\right)^{\gamma(-1)-1}e^{-\frac{Z_T r}{a_0}}
\label{rbwf}
\end{eqnarray}
with the Bohr radius denoted by $a_0$.
For the continuum we use the radial functions 
\begin{eqnarray}
\left(
\begin{array}{c}
g_{E,\kappa}(r)\\ f_{E,\kappa}(r)
\end{array}\right) & = & \left(\frac{E+m_e}{E-m_e}\right)^{\frac 1 4} \frac k
{\pi^\frac 1 2} N\, (kr)^{\gamma(\kappa)-1}\nonumber \\
& & \times\left(
\begin{array}{c}
\mbox{Re}\\ \mbox{sgn}(E)\sqrt{\frac{E-m_e}{E+m_e}}\ \mbox{Im}
\end{array}\right)\left[e^{-i(kr+\phi)}\,_1F_1(\gamma(\kappa)+i\eta,
2\gamma(\kappa)+1,2ikr)\right]
\label{rcwf}
\end{eqnarray}
which are normalized according to
\begin{equation}
\int_0^\infty dr\ r^2\,[g_{E,\kappa}\,g_{E',\kappa}
+f_{E,\kappa}\,f_{E',\kappa}]=\delta(E-E')\ \ .
\label{nm}
\end{equation}
In (\ref{rbwf}) and (\ref{rcwf}) $\gamma(\kappa)$, 
$k$, $\eta$ and $N$ are given by
\begin{equation}
\gamma(\kappa)=\sqrt{\kappa^2-\zeta^2},\ \ \ 
k=\sqrt{E^2-m_e^2},\ \ \ 
\eta=\frac{\zeta E}{k},\ \ \ 
N=\frac{2^{\gamma(\kappa)} e^{\frac{\pi \eta} 2}|\Gamma(\gamma(\kappa)
+1+i\eta)|}{\Gamma(2\gamma(\kappa)+1)}\ \ .
\end{equation}
We rewrite the expression in parenthesis in the matrix element of 
(\ref{cscg}) in the spherical
basis $(\vec e_0,\vec e_{\pm 1})$ with $ \vec e_0=\frac{\vec q} s$ and 
use the expansion of the vector plane waves into electromagnetic 
multipoles. From the orthogonality of the spherical harmonics we get the 
incoherent sum over the multipoles. After some algebra one obtains a 
relatively simple expression for the cross section. 
\begin{equation}
\sigma_{tot}^K=32\pi^2\left(\frac{Z_P \alpha}{\beta}\right)^2
\int \limits_{m_e}^\infty dE_i \int \limits_{q_z}^\infty
ds \left\{\frac{T_l}{s^3}+\frac{\beta^2}{2}\frac{s}{[s^2-(\beta q_z)^2]^2}
\left(1-\frac{q_z^2}{s^2}\right)T_\perp\right\}\ ,
\label{sk}
\end{equation}
with $T_l$ and $T_\perp$ given by
\begin{eqnarray}
T_l & = & \sum_{\kappa_i,l}\frac{(2j_i+1)(2j_f+1)}{4\pi}(2l+1)\frac 1 2 
\left[1+(-1)^{l_f+l+l_i}\right]\left|J^l(E_i,s)\right|^2
\left(
\begin{array}{ccc}
j_f & l & j_i \\
\frac 1 2 & 0 & -\frac 1 2
\end{array}
\right)^2\ \ ,\\
T_{\perp} & = & \sum_{\kappa_i,l}\frac{(2j_i+1)(2j_f+1)}{4\pi}(2l+1)
\Biggl\{
\frac 1 2 \left[1-(-1)^{l_f+l+l_i}\right]\frac 1 {l(l+1)}
\left|(\kappa_f+\kappa_i)\,I_l^+( E_i,s)\right|^2\nonumber \\ 
& & +\frac{1}{2}\left[1+(-1)^{l_f+l+l_i}\right]\frac 1 {(2l+1)^2}
\Biggl|\left(\frac{l+1}{l}
\right)^\frac 1 2 \left[(\kappa_f-\kappa_i)\,I_{l-1}^+( E_i,s)
-l\ I_{l-1}^-( E_i,s)\right]\nonumber \\ 
& & -\left(\frac{l}{l+1}\right)^\frac 1 2 \left[(\kappa_f-\kappa_i)
\,I_{l+1}^+( E_i,s)+(l+1)\,I_{l+1}^-( E_i,s)\right]\Biggr|^2
\Biggr\}
\left(
\begin{array}{ccc} 
j_f & l & j_i \\
\frac 1 2 & 0 & -\frac 1 2
\end{array}
\right)^2\ \ .
\label{Tperpfunc}
\end{eqnarray}
The radial integrals are given by
\begin{eqnarray}
J^l(E_i,s) & = & \int \limits_0^\infty dr\,r^2 j_l(sr)[g_{\kappa_f}(r)
g_{E_i,\kappa_i}(r)+f_{\kappa_f}(r)f_{E_i,\kappa_i}(r)]\ \ ,\\
I_l^\pm (E_i,s) & = & \int \limits_0^\infty dr\,r^2 j_l(sr)[g_{\kappa_f}(r)
f_{E_i,\kappa_i}(r)\pm f_{\kappa_f}(r)g_{E_i,\kappa_i}(r)]
\label{formfactors}
\end{eqnarray}
and are evaluated quickly by the method presented in \cite{dirk}.
\section{Results}
We focus our attention on the process $\overline p + Z_P \rightarrow
\overline H(1s)+Z_P+e^-$.  
In Fig. \ref{cross} and Table \ref{tab} 
the total cross section for $Z_P=Z_T=1$ 
is given as a function of the Lorentz $\gamma$-factor of the projectile 
in the target rest frame. The cross sections are calculated 
numerically with (\ref{sk}) - (\ref{formfactors}).
\begin{figure}[h]
\caption{The solid line shows the total cross section $\sigma_{tot}^K$ 
for antihydrogen production with a positron bound in the 
{\em K}-shell ($Z_P=Z_T=1$) as a function of the Lorentz $\gamma$-factor of the 
projectile in the target rest frame. The dotted line presents the part of 
(12) containing $T_\perp$. The cross section scales with $Z_P^2$.}
\label{cross}
\end{figure}
\begin{table} 
\caption{The total cross section for antihydrogen 
production ($Z_P=Z_T=1$) is given for different $\gamma$'s of the 
projectile in the target rest frame for the Fermilab experiment 
\protect\cite{fermilab}. 
The {\em K}-shell capture cross section is 
denoted by $\overline H(1s)$ and the cross section for 
capture into all shells  by $\overline H(all)$. The contribution of all higher
shells is estimated to 20\% of the {\em K} capture 
\protect\cite{agger,bertulani}.}
\label{tab}
\begin{tabular}{lcc}
$\gamma_{target\ rest\ frame}$ & $\sigma_{\overline H(1s)}$ [pb] & 
$\sigma_{\overline H(all)}$ [pb] \\
\hline
5&$5.73*10^{-1}$&$6.88*10^{-1}$\\
6&$7.90*10^{-1}$&$9.48*10^{-1}$\\
7&$1.00$&$1.20$\\
\end{tabular}
\end{table}
\widetext
We compare our results with experimental data measured 
in the low energy region for La$^{57+}$ target \cite{meyerhof}.
Our calculations for {\em K}-shell bound-free pair production of 
La$^{57+}$ with projectile energies 0.405 GeV/u ($\gamma =1.43$), 
0.956 GeV/u ($\gamma =2.026$), and 1.3 GeV/u ($\gamma =2.40$) yields
$\sigma^K/Z_P^2=$ $9.93*10^{-7}$b, $9.12*10^{-6}$b, 
and $1.78*10^{-5}$b, respectively.
Adding contribution from higher shells, which are assumed to 
be about 20\% \cite{agger,bertulani},
we find good agreement with the experimental results.
We also compare our results with those of Becker \cite{becker}.
Our results agree well with the results given there for 1 GeV amu$^{-1}$ 
($\gamma =2.07$) and 15 GeV amu$^{-1}$ ($\gamma=16.1$). Our results 
for $Z_P=Z_T=1$ are
$\sigma_{tot}^K=3.92*10^{-2}$pb and $\sigma_{tot}^K=2.55$pb, respectively.
\section{Correspondence to EPA}
To compare (\ref{sk}) with the photoproduction cross section we rewrite 
it as:
\begin{equation}
\sigma_{tot}^K=16\pi^2\left(\frac{Z_P\alpha}{\beta}\right)^2
\int \limits_{m_e+E_f}^\infty d\omega \int \limits_0^\infty
d(q_\perp^2)\frac 1 { q_\perp^2+\left(\frac\omega \beta \right)^2} 
\left\{\frac{1}{q_\perp^2+\left(\frac\omega \beta\right)^2} 
T_l +\frac{\beta^2}{2}\frac{q_\perp^2}{\left[q_\perp^2+
\left(\frac\omega {\beta\gamma}\right)^2\right]^2}
T_\perp\right\}
\label{br}
\end{equation}
with $\omega=E_f-E_i$ and $q_{\perp}^2=s^2-q_z^2$; 
$q_z=\frac \omega \beta$.
$\sigma_{tot}^K$ as a function of 
$\omega$ is shown in Fig. \ref{omega}.
\begin{figure}[h]
\caption{The cross section for antihydrogen 
production as a function of $\omega$. The solid line 
shows the cross section for the Lorentz factor $\gamma=3$ of the projectile 
in the target rest frame. The dashed line shows the cross section for 
$\gamma=6$ and the dotted line for $\gamma=10$.}
\label{omega}
\end{figure}
The S-matrix element for photo-induced bound-free pair production 
(also known as photo-induced {\em K}-shell capture) in Coulomb gauge is 
given by  
\begin{equation}
S_{fi}=-ie\int_{-\infty}^{\infty}dt\left\langle
\psi_f(\vec r)|\vec {\alpha} \vec e_\mu e^{i\vec k \vec r}|\psi_i(\vec r)
\right\rangle
e^{i\left(E_f-E_i-\omega\right)t}\ \ .
\end{equation}
Using the multipole expansion for $\vec e_\mu e^{i\vec k \vec r}$ 
we find the expression for the total cross section:
\begin{equation}
\sigma_\gamma^{K}(\omega)=
\frac{8\pi^3\alpha}{\omega}T_\perp(\omega,q^2=0)
\label{photo}\ \ ,
\end{equation}
where $T_\perp$ is the same as in (\ref{Tperpfunc}), with $s=\omega$. 
Here we denote the four-momentum of the photon by $q$.
If in (\ref{br}) $T_l$ is omitted, we can define 
the photo-induced cross section for 'virtual' photons
(see \cite{brodsky} Eq. (1)) as
\begin{equation}
\sigma_{\gamma^*}^{K}(\omega,q^2)=8\pi^3\alpha
\frac \omega
{q_\perp^2+\left(\frac\omega \beta\right)^2}T_\perp(\omega,q^2)\ \ .
\label{photovirtual}
\end{equation} 
The expression (\ref{br}) can now be written as 
$\sigma_{tot}^K(\sigma_{\gamma^*})$. 
At this stage we can introduce the equivalent photon approximation (EPA).
In this approximation the $q^2$-dependence of $\sigma_{\gamma^*}^K$ 
is neglected and (\ref{photovirtual}) is replaced by the cross section 
for real photons (\ref{photo}). Now the integral over $q_\perp$ would diverge
and we must introduce a suitable cutoff. Now we can write the cross section in 
the equivalent photon approximation as 
\begin{equation}
\sigma_{EPA}^K= \frac{Z_P^2\alpha}{\pi}
\int \limits_{m_e+E_f}^{\infty} d\omega \int 
\limits_0^{q_{\perp max}^2}d(q_\perp^2)\frac 1 \omega\ 
\frac{q_\perp^2}{\left[q_\perp^2+\left(\frac\omega 
{\beta\gamma}\right)^2\right]^2}\ 
\sigma_{\gamma}^K(\omega)\ \ .
\label{epa} 
\end{equation}
$T_\perp(\omega,q^2)$ as a function of the Lorentz-invariant variable
$Q^2=-q^2$ and $\omega$ is given in Fig. \ref{Tperp}.
The momentum $q_\perp^2$ is related to $q^2$ by
\begin{equation}
q^2=\omega^2-q_z^2-q_\perp^2=-\left(\frac\omega{\beta\gamma}
\right)^2-q_\perp^2\ \ .
\label{q2}
\end{equation} 
\begin{figure}[h]
\caption{$T_{\perp}$ as a function of $Q^2=-q^2$ for fixed 
$\omega$-values. The solid line shows $T_\perp$ 
for $\omega=2.1$, the dashed line for $\omega=3$, the 
dotted line for $\omega=5$ and the dash-dotted line for $\omega=10$.}
\label{Tperp}
\end{figure}
\begin{figure}[h]
\caption{$T_l$ as a function of $Q^2$ for fixed 
$\omega$-values. The solid line shows $T_l$ 
for $\omega=2.1$, the dashed line for $\omega=3$, the 
dotted line for $\omega=5$ and the dash-dotted line for $\omega=10$.}
\label{Tlong}
\end{figure}
From Fig. \ref{Tperp} one can extract the cutoff parameter
for the EPA calculation. 
For the range of $\omega$-values contributing significantly to the
total cross section (see Fig. \ref{omega})
$T_\perp$ is almost constant up to 
$Q^2_{max}\approx 4 m_e^2$. Beyond this value it falls off rapidly.
Therefore $q^2_{\perp max}$ is given by
$q^2_{max}=Q^2_{max}-(\frac \omega {v\gamma})^2$. 
In Fig. \ref{Tlong} we also show $T_l(\omega,q^2)$ as a function of 
$Q^2=-q^2$ and $\omega$ which has a similar behavior as $T_\perp$.
In Fig. \ref{epafig} we compare EPA results with the SCA calculation 
(\ref{sk}).
\begin{figure}[h]
\caption{The solid line shows the full SCA calculation results. The dotted line
shows EPA results with $q^2_{\perp max}=4m_e^2-(\frac \omega {v\gamma})^2$. 
For the photo-induced cross section in EPA calculation 
we used the expression (\protect\ref{achi}).}
\label{epafig}
\end{figure}
In the high energy region EPA fits correctly the total cross section. 
For $\gamma>>1$, the part of the integral for the total cross section 
in (\ref{sk}) containing $T_l$ tends to a  constant (see Fig. \ref{const}).
Therefore for high energies of the colliding 
particles $\sigma_{tot}=\sigma_{EPA}+\sigma_{const}$. 
For high $\gamma$-values we are in agreement with the EPA results 
of \cite{aste,agger}.
Due to the $\omega$-dependence of the cutoff $q^2_{\perp max}=
4m_e^2-(\frac \omega {v\gamma})^2$ the EPA formula is valid 
down to $\gamma=3$ . But in the low energy region ($\gamma<3$) EPA fails. 
Further for small $\gamma$'s the contribution of $T_l$ to the total 
cross section can no longer be neglected but has been ignored in the 
derivation of the EPA formula (\ref{epa}).
\begin{figure}[h]
\caption{The solid line presents the part of (12) containing $T_l$.}
\label{const}
\end{figure}

An analytical expression for the photo cross section (\ref{photo})
can be obtained from Sauter's formula for the photoelectric effect 
\cite{sauter} by means of crossing symmetry:
\begin{equation}
\sigma_\gamma^{K}=4\pi\alpha^6\frac{Z_T^5}{m_e^2}\left[
\frac{k_f}{(1-\varepsilon_i)^4}\left(\varepsilon_i^2-\frac{2}{3}\varepsilon_i
+\frac 4 3 -\frac{2-\varepsilon_i}{k_i}\ln(k_i-\varepsilon_i)
\right)\right]\ ,
\label{achi}
\end{equation}
with $\varepsilon_i=\frac{E_i}{m_e}$ and $k_i=\sqrt{\varepsilon_i^2-1}$.  
It is an approximation for $\alpha Z_T<<1$ and relativistic velocities of 
the unbound electron. We compare the exact cross section (\ref{photo})
with the analytic result (\ref{achi}).
Very good agreement is expected for $Z_T=1$. This is shown in 
Fig. \ref{achicheck}. With (\ref{photo}) it is also possible to calculate 
$\sigma_\gamma$ for $\alpha Z_T \sim 1$ where (\ref{achi}) overestimates 
the cross section (see \cite{aste,johnson}).
\begin{figure}[h]
\caption{The solid line gives the results of the exact calculations  
(\protect\ref{photo}) and the dotted line the approximated 
analytical expression (\protect\ref{achi}) for $Z_P=Z_T=1$.}
\label{achicheck}
\end{figure}
\section{Conclusion}
We calculate bound-free pair production in the semiclassical 
approximation (SCA) using exact Dirac-Coulomb wave functions. 
We compare results for ion-ion collision with 
experimental and theoretical data of \cite{aste,agger,meyerhof,becker}
and find good agreement with their results.

Our calculations have the advantage remaining valid also when the condition
$Z\alpha<<1$ is not fulfilled, since exact Dirac wave functions are used. 
We give expressions for the cross sections already integrated over the 
angular variables of the free electron (or positron).
The radial form factor integrals are evaluated quickly by means of 
recurrence relations as given in \cite{dirk}. Therefore we are able 
to sum over many partial waves and this allows the extension of our 
calculations up to high values of $\gamma$.

We compare the exact results which those of the equivalent photon 
approximation (EPA).
We show that good agreement can be obtained already starting at 
$\gamma\geq 3$ if one uses an $\omega$-dependent
cutoff of $q_{\perp max}=\sqrt{Q_{max}^2-(\frac{\omega}{\beta \gamma})^2}$
with $Q_{max}^2\approx 4m_e^2$. This justifies the cutoff chosen in 
\cite{aste,agger} at high $\gamma$-values.
At high energies the contribution of the longitudinal photon tends to a
constant. Therefore the total cross section is found to be of the form 
$\sigma_{tot}=\sigma_{EPA}+\sigma_{const}$. 


\begin{references}
\bibitem{cern}PS210 collaboration, W. Oelert, spokesperson, G. Baur et al:
Phys. Lett. B 368 (1996) 251
\bibitem{datz}S. Datz et al: Nucl. and Instr. Meth. in Phys. Res. 124 B 
(1997) 129
\bibitem{brodsky}C. T. Munger, S. J. Brodsky, I. Schmidt:
Physical Review D 49 (1994) 3228
\bibitem{aste}A. Aste, K. Hencken, D. Trautmann, G. Baur:
Phys. Rev. A 50 (1994) 3980 
\bibitem{agger}C. K. Agger, A. H. So$\hspace{-.5em}/$rensen:
Phys. Rev. A 55 (1997) 402
\bibitem{baur}G. Baur: Phys. Lett. B 311 (1993) 2002
\bibitem{baltz}A. J. Baltz: Phys. Rev. Lett. 78 (1997) 1231
\bibitem{eichler}J. Eichler, W. E. Meyerhof:
Relativistic Atomic Collisions: Academic Press (1995); J. Eichler:
Theory Of Relativistic Ion-Atom Collisons: Physics Reports 193, 
Nos. 4 \& 5 (1990) 165
\bibitem{amundsen}P. A. Amundsen, K. Aashamar: J. Phys. B 14 (1981) 4047
\bibitem{rose}M. E. Rose: Relativistic Electron Theory: Wiley, New York (1961)
\bibitem{dirk} D. Trautmann, G. Baur, F. R\"osel: 
J. Phys. B 16 (1983) 3005
\bibitem{fermilab}M. Mandelkern: letter of intent, experiment E862, Fermilab,
: available at http://fnphyx-fnal.gov/e862/e862.html 
\bibitem{bertulani}C. A. Bertulani, G. Baur:
Phys. Rep. 163 (1988) 299
\bibitem{meyerhof} A. Belkacem et al: Phys. Rev. Lett. 73 (1994) 2432 
\bibitem{becker}U. Becker: J. Phys. B 20 (1987) 6563
\bibitem{sauter}F. Sauter: Ann. Phys. (Leipzig) 11 (1931) 454
\bibitem{johnson} W. R. Johnson, D. J. Buss, C. O. Carroll:
Phys. Rev A 135 (1964) 1232
\end{references}
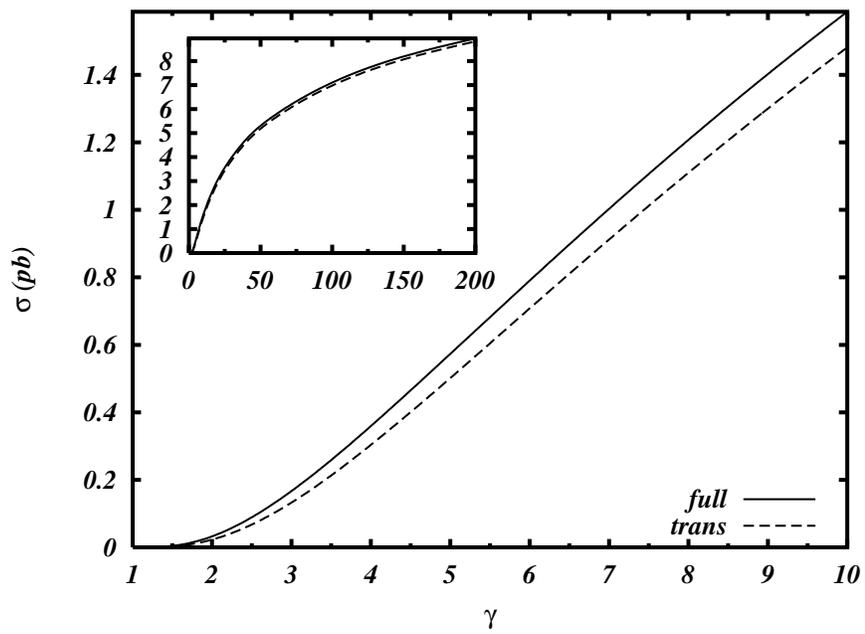
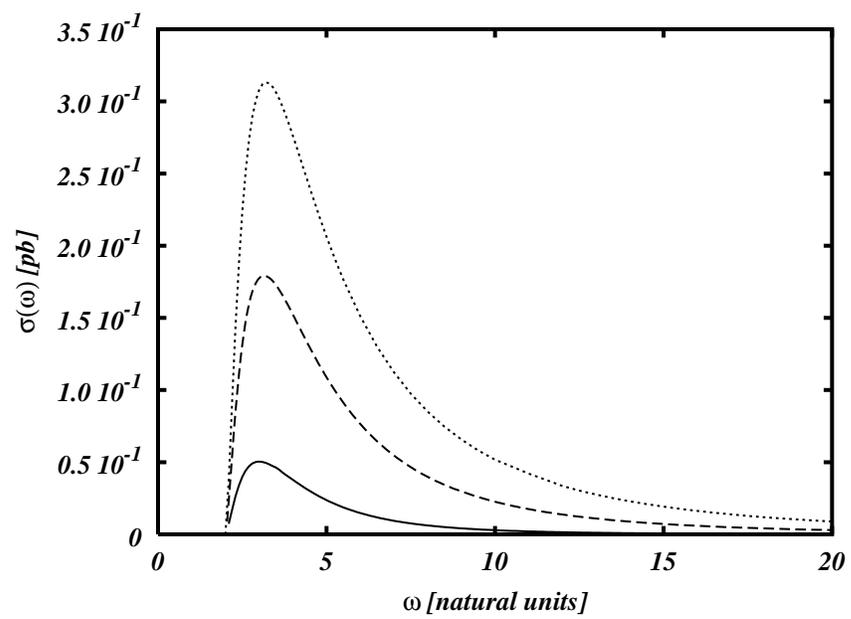
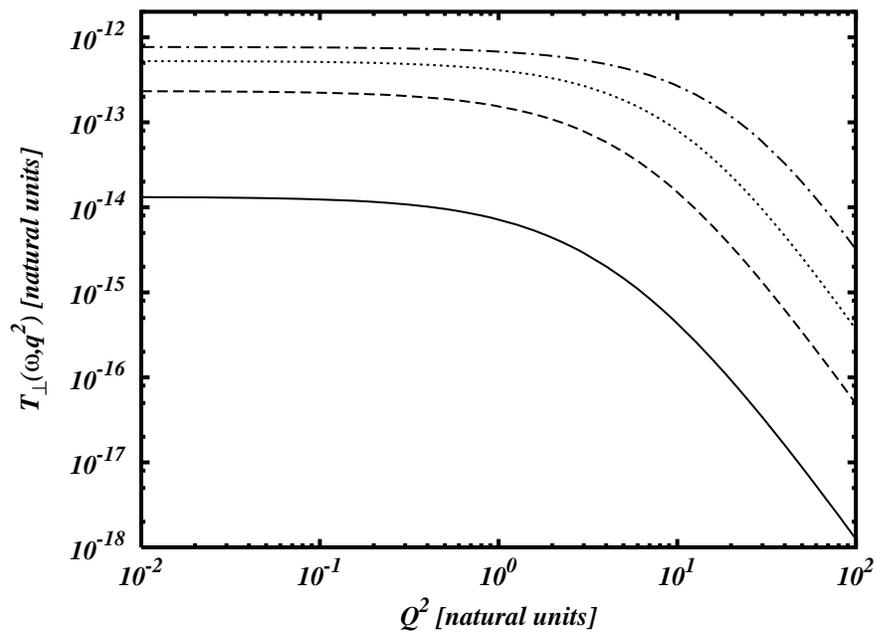
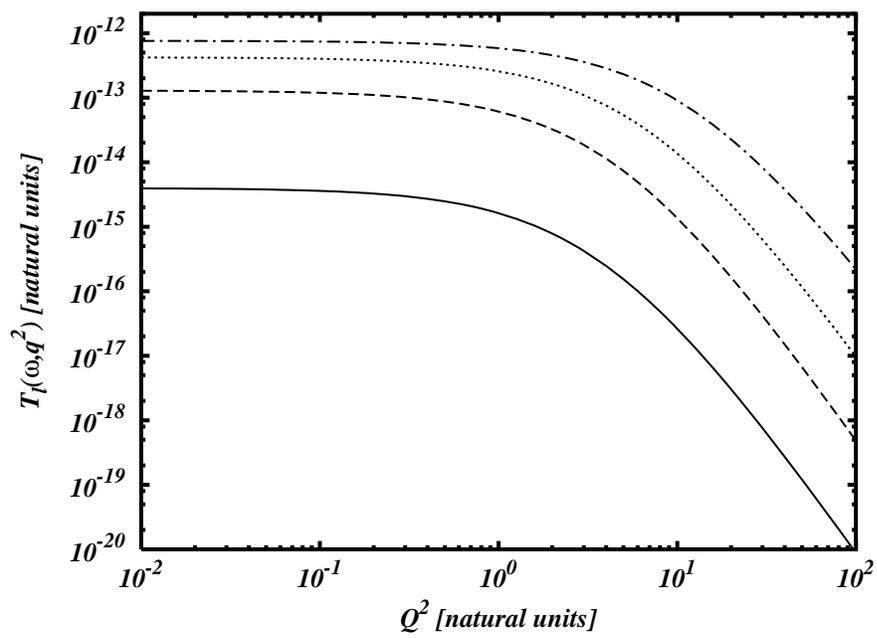
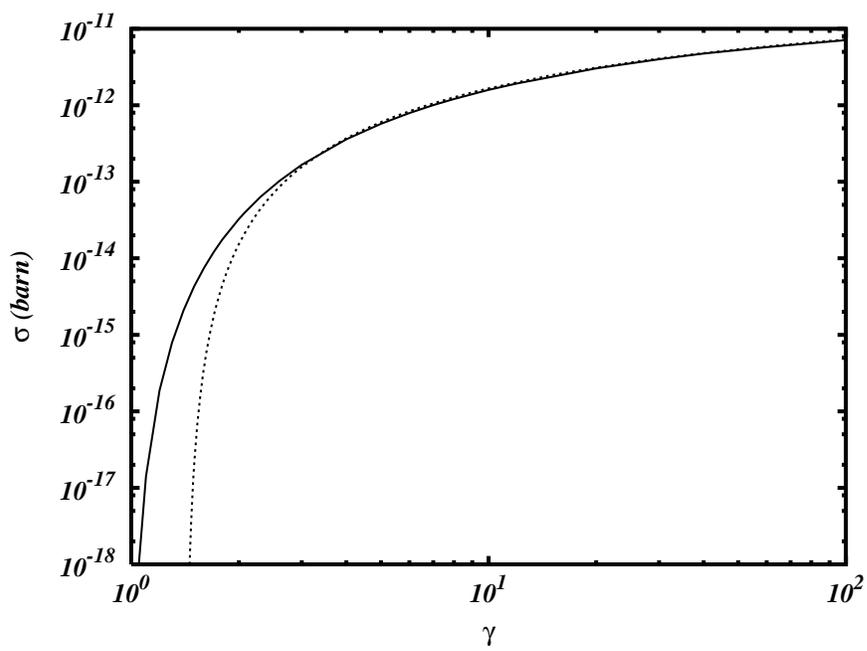
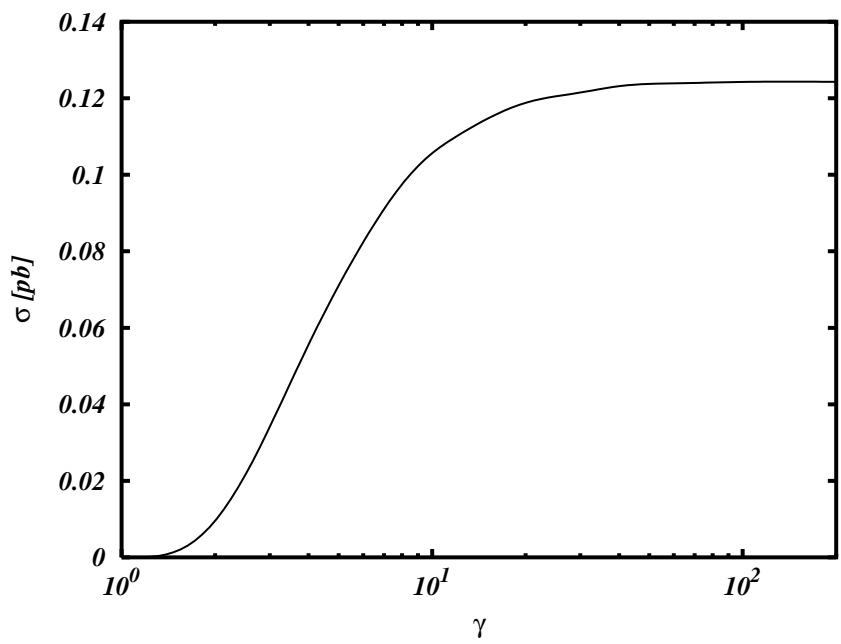
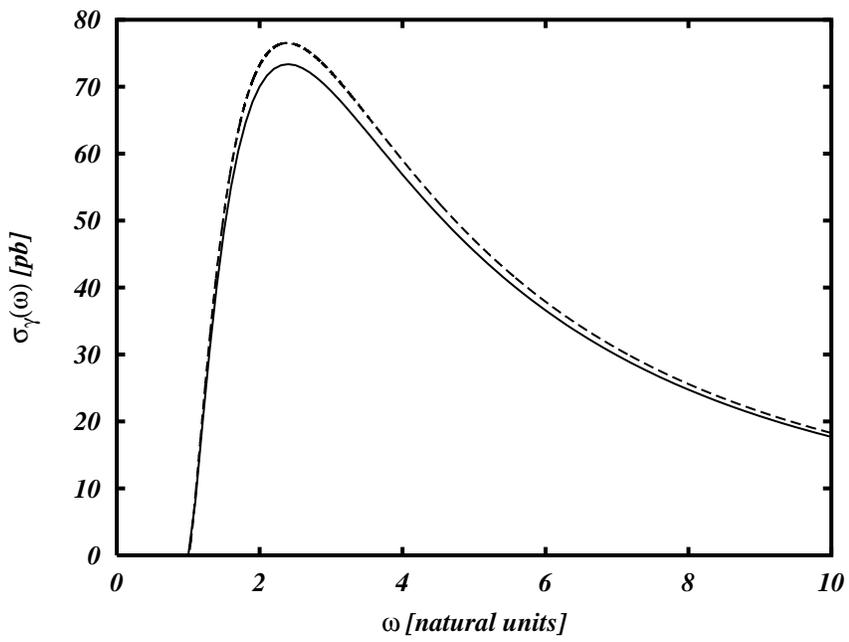
\end{document}